\documentclass[doublecol]{epl2}

\usepackage{amsmath, amsthm}
\usepackage{amsthm}
\usepackage{float}
\usepackage[dvipsnames, svgnames, x11names]{xcolor}
\usepackage{graphicx} 
\usepackage{epstopdf}
\usepackage{lipsum}
\usepackage{caption}
\usepackage{pifont}
\usepackage{amssymb}
\usepackage{algorithm }
\usepackage{algorithmic}
\usepackage{amsfonts}

\def\qed{\leavevmode\unskip\penalty9999 \hbox{}\nobreak\hfill
     \quad\hbox{\leavevmode  \hbox to.77778em{%
               \hfil\vrule   \vbox to.675em%
               {\hrule width.6em\vfil\hrule}\vrule\hfil}}
     \par\vskip3pt}

\newtheorem{remark}{Remark}

\newtheorem{theorem}{Theorem}
\newtheorem{corollary}{Corollary}
\newtheorem{lemma}{Lemma}
\newtheorem{example}{Example}

\title{Improved tests of genuine entanglement for multiqudits}
\shorttitle{Improved tests of genuine tripartite entanglement} 

\author{Xia Zhang\inst{1} \and Naihuan Jing\inst{2} \and Hui Zhao\inst{3} \and Ming Liu\inst{4} \and Haitao Ma\inst{5}}
\shortauthor{X. Zhang \etal}

\institute{
  \inst{1} Guangdong Baiyun University, Guangzhou 510450, China\\
  \inst{2} North Carolina State University, Raleigh, NC 27695, USA\\
  \inst{3} Beijing University of Technology, Beijing 100124, China\\
  \inst{4} South China University of Tecchnology, Guangzhou 510640, China\\
  \inst{5} Harbin Engineering University, Harbin 15001, China
}

\abstract{
We give an improved criterion of genuine multipartite entanglement for an important class of
multipartite quantum states using generalized Bloch representations of
 the density matrices. The practical criterion is designed based on the Weyl operators and can be used for detecting genuine multipartite entanglement
 in higher dimensional systems. The test is shown to be significantly stronger than
some of the most recent criteria.}

\begin{document}

\maketitle

\section{Introduction}
Quantum entanglement is an important phenomenon in quantum systems responsible for quantum superiority.
Many applications of quantum entanglement have been found, for instance, in entanglement swapping \cite{bvk}, quantum cryptography \cite{eak} and quantum secure communication \cite{bw}.

The genuine multipartite entanglement (GME) is perhaps one of the most significant quantum phenomenon \cite{hp,tg} and the study of measuring
GME has been a nontrivial problem in quantum information \cite{pa,hhh,mw,nki,u,ss,bjl,pv,hong,lrb,skt}. Many criteria were found for tripartite states: separability of bipartite quantum systems via Bloch representation \cite{Vicente}, sufficient tests \cite{akb} in the vicinity of the GHZ state, the W states and the PPT entangled states,sufficient conditions for three-particle entanglement\cite{su},classification of mixed three-qubit state\cite{abl}. Similar criteria were found using local sum uncertainty relations \cite{ymm} as well as for genuine tripartite entanglement based on partial transposition and realignment of density matrices \cite{mes}. Criterion for tripartite entanglement was also studied \cite{ysc} in terms of quantum Fisher information. Genuine multipartite entanglement state was discussed in an electron-positron system in \cite{pa}. In \cite{va}, the authors detected GME based on local harnessing or dephasing quantum non-Markovian operations. Tests based on norms of correlation tensors were given in \cite{lmj,dgh}. For tripartite and four-partite states, separability tests were
 given by using the Bloch representation \cite{zzj}, by the matrix method \cite{zljw}, by the upper bound of the Bloch vectors \cite{lww},and by using witness operators\cite{bek}. Detection of genuine tripartite entanglement was given by multiple sequential observers\cite{mdg},and by  local marginals\cite{bpr}. For higher dimensional quantum system, the separable criteria and $k$-separable criteria for general $n$-partite quantum states were also presented in \cite{hgy,lwf,xzz}.

Block representation has been used in detecting bipartite entanglement widely, see for example \cite{Vicente,ref35,ref36,ref37,ref38,ref39}.
In this paper, we will use the generalized Bloch representation and the Weyl operators to study the genuine entanglement of multipartite quantum systems.
The Weyl representation uses a uniformed generators of the Lie algebra $\mathfrak{su}(d)$ to generate the principal basis, thus it is
easier to treat higher dimensional cases in practical computation.
The tests are presented by using some special matrices from the density matrices, and we derive a better mathematical bound
for the norm.
Our new test has improved
some of the recently known tests obtained by similar or different consideration.

The layout of the paper is as follows. In section 2, we quickly review some basic notions concerning the Weyl operators as generalized Pauli spin matrices
and discuss their main properties. We then construct certain tensors based on the density matrices of the quantum system to
estimate its bound under various situations, which naturally lead to new criterion for the GME. In section 3, we discuss how to generalize our results to higher dimensional cases, and the conclusion is given in section 4. Our test is found to be significantly stronger than some of the recently available criteria when
the number of particles is large.

\section{Genuine tripartite entanglement}
We start by considering the GME for tripartite states.
Let $E_{ij}$ be the unit matrices of size $d$, where $(E_{ij})_{kl}=\delta_{ki}\delta_{jl}$. Let $\omega$ be a fixed $d$-th primitive root of unity, then the principal basis matrices or the Weyl operators are defined by
\begin{equation}\label{1}
  A_{ij}=\sum\limits_{m\in \mathbb Z_{d}}\omega^{im}E_{m,m+j},
\end{equation}
summed over $\mathbb Z_{d}=\mathbb Z/d\mathbb Z$. The Weyl operators obey the rule \cite{bgj, hjz}:
$
  A_{ij}A_{kl}=\omega^{jk}A_{i+k,j+l},
$
and $A_{00}=I$, so $A_{i,j}^{\dagger}=\omega^{ij}A_{-i,-j}$ and $tr(A_{ij}A_{kl}^{\dagger})=\delta_{ik}\delta_{jl}d$.
For a column vector $X$ let $\|X\|=\sqrt{X^{\dagger}\cdot X}$ be the norm, and we also use it for the norm of a square matrix viewed as a
prolonged column vector. For $A\in \mathbb{C}^{m\times n}$ let $\|A\|_{tr}=\sum\sigma_i=tr\sqrt{AA^\dagger}$,
where $\sigma_i$ are the singular values of $A$.

Any state on $H^{d}$ can be written uniquely as
\begin{equation}\label{e:bloch}
\rho=\frac{1}{d}(I_{d}+\sum\limits_{(i_1,j_1)\atop\neq(0,0)}u_{i_1j_1}A_{i_1j_1})=\frac{1}{d}(I_{d}+T\cdot A),
\end{equation}
summed over nonzero pairs of modulo $d$ integers, where $A$ is the $(d^2-1)$-vector of nonzero Weyl operators,
$T$ is the column vector of size $d^2-1$ with entries $u_{i_1j_1}=tr(\rho(A_{i_1j_1})^{\dagger})$, and $T\cdot A$ is the (complex) dot product in $\mathbb C^{d^2-1}$.

\begin{lemma}\label{lemma:1} For any state $\rho\in H^{d}$, we have
$\|T\|^2\leq d-1$.
\end{lemma}
{\it Proof}~~Since $tr(\rho^2)\leq1$, we have $tr(\rho^2)=tr(\rho\rho^{\dagger})=\frac{1}{d}(1+\|T\|^2)\leq1,$
therefore $\|T\|^2\leq d-1$.
\qed

For any state $\rho\in H_1^{d_1}\otimes H_2^{d_2}$,
$\rho$ has the following Bloch-like expression based on the Weyl operators:
\begin{equation}\label{3}
\begin{split}
\rho&=\frac{1}{d_1d_2}(I_{d_1d_2}+T^{(1)}\cdot A_{d_1}\otimes I_{d_2}\\
&+I_{d_1}\otimes T^{(2)}\cdot A_{d_2}+T^{(12)}\cdot(A_{d_1}\otimes A_{d_2}))
\end{split}
\end{equation}
where $T^{(1)}$, $T^{(2)}$, $T^{(12)}$ are the matrices with entries $u_{i_1j_1}^{(1)}=tr(\rho(A_{i_1j_1}^{(1)})^{\dagger}\otimes I_{d_2})$, $u_{i_1j_1}^{(2)}=tr(\rho I_{d_1}\otimes(A_{i_2j_2}^{(2)})^{\dagger})$, $v_{i_1j_1,i_2j_2}^{(12)}=tr(\rho(A_{i_1j_1}^{(1)})^{\dagger}\otimes(A_{i_2j_2}^{(2)})^{\dagger})$, and
$A_{d_s}=(A_{i_sj_s})$ 
is the column vector of Weyl operators on $H_{d_s}$ ($s=1, 2$). 

Setting $\mathfrak{m}_{fg}=\min\{d_fd_g-\frac{d_f}{d_g},d_fd_g-\frac{d_g}{d_f}\}$, we have the following lemma.
\begin{lemma}\label{2}
Let $\rho\in H_1^{d_1}\otimes H_2^{d_2}$ be a quantum state, we have $\|T^{(12)}\|^2\leq \mathfrak{m}_{12}$.
\end{lemma}
{\it Proof}~~It suffice to show for a pure state $\rho$, then $tr(\rho^2)=1$ and
\begin{equation}\label{4}
\begin{split}
tr(\rho^2)& 
=\frac{1}{d_1d_2}(1+\|T^{(1)}\|^2+\|T^{(2)}\|^2+\|T^{(12)}\|^2).
\end{split}
\end{equation}
Note that $tr(\rho_{H_1}^2)=tr(\rho_{H_2}^2)$, then $\frac{1}{d_1}(1+\|T^{(1)}\|^2)=\frac{1}{d_2}(1+\|T^{(2)}\|^2)$ so
\begin{align*}
&\frac{1}{d_1^2}(1+\|T^{(1)}\|^2)+\frac{1}{d_2^2}(1+\|T^{(2)}\|^2)\\
&=
\frac{1}{d_1d_2}(2+\|T^{(1)}\|^2+\|T^{(2)}\|^2).
\end{align*}
Therefore,
\begin{equation}\label{5}
\begin{split}
&\|T^{(12)}\|^2=d_1d_2-1-\|T^{(1)}\|^2-\|T^{(2)}\|^2\\
&=d_1d_2-1-[\frac{d_1+d_2}{d_2}(1+\|T^{(2)}\|^2)-2]\\
&=d_1d_2-1-[\frac{d_1+d_2}{d_1}(1+\|T^{(1)}\|^2)-2]
\leq \mathfrak{m}_{12}
\end{split}
\end{equation}
\qed

To represent a tripartite state, we introduce some notations. Let $A_{(i)}$ be the column vector whose entries are
nonidentity Weyl operators acting on the $i$th factor of
the space $H_{d_1}\otimes H_{d_2}\otimes H_{d_3}$, i.e. the general entries of $A_{(2)}$ are of the form $I_{d_1}\otimes A_{ij}\otimes I_{d_3}$. Then the matrix of Weyl operators whose entries act on $(i, j)$-factor
of $H_{d_1}\otimes H_{d_2}\otimes H_{d_3}$ can be written as $A_{(i)}A_{(j)}$, for example, the general entries of $A_{(1)}A_{(3)}$ are of the form $A_{i_1j_1}\otimes I_{d_2}\otimes A_{i_2j_2}$.
Then a general tripartite state $\rho\in H_{1}^{d_1}\otimes H_{2}^{d_2}\otimes H_{3}^{d_3}$ can be written uniquely as follows.
\begin{equation}\label{6}
\begin{split}
\rho
&=\frac{1}{d_1d_2d_3}(I_{d_1d_2d_3}+\sum_{1\leq i\leq 3}T^{(i)}\cdot A_{(i)}\\
&+\sum_{1\leq i<j\leq 3}T^{(ij)}\cdot A_{(ij)}+T^{(123)}\cdot A_{(123)})
\end{split}
\end{equation}
where $T^{(i)}=(u_{a_ib_i}^{(i)})$, $T^{(ij)}=(v_{a_ib_i,a_jb_j}^{(ij)})$ and $T^{(123)}=(r_{ij,kl,st}) $ with $u_{a_ib_i}^{(i)}=tr(\rho (A_{a_ib_i}^{(i)})^{\dagger}\otimes I\otimes I)$, $v_{a_ib_i,a_jb_j}^{(ij)}=tr(\rho (A_{a_ib_i}^{(i)})^{\dagger}\otimes (A_{a_jb_j}^{(j)})^{\dagger}\otimes I)$, and $r_{ij,kl,st}=tr(\rho (A_{ij}^{(1)})^{\dagger}\otimes (A_{kl}^{(2)})^{\dagger}\otimes (A_{st}^{(3)})^{\dagger})$.

\medskip

We now define some useful block matrices out of the correlation tensor of $\rho$. For real numbers $\alpha$, $\beta$, ${\gamma}$ and distinct indices
$i, j, k\in\{1,2,3\}$, set
\begin{equation}\label{7}
N^{i|jk}=\alpha S_0^{i|j}+\beta S^{i|k}+\gamma S^{i|jk},
\end{equation}
where $S_0^{i|j}=[S^{i|j}~~O_i]$ is a block matrix with submatrix $S^{i|j}=[v_{a_ib_i,a_jb_j}^{(ij)}]$ of size $(d_i^2-1)\times(d_j^2-1)$ and $O_i$ being the zero matrix of size $(d_i^2-1)\times(d_j^2-1)(d_k^2-2)$ and $S^{i|jk}=[r_{ab,cd,ef}]$ is a $(d_i^2-1)\times(d_j^2-1)(d_k^2-1)$ matrix. For example, when $\rho\in H_1^2\otimes H_2^2\otimes H_3^3$, $N^{2|13}=\alpha S_0^{2|1}+  \beta S^{2|3}  +\gamma S^{2|13}$, where
$$S^{2|1}=\left[
  \begin{array}{ccc}
    v_{01,01}^{(21)} & v_{10,01}^{(21)} & v_{11,01}^{(21)} \\
    v_{01,10}^{(21)} & v_{10,10}^{(21)} & v_{11,10}^{(21)} \\
    v_{01,11}^{(21)} & v_{10,11}^{(21)} & v_{11,11}^{(21)} \\
  \end{array}
\right],
$$
$$
S^{2|13}=\left[
  \begin{array}{ccccc}
    r_{01,01,01}  & \cdots & r_{01,01,22} & \cdots & r_{11,01,22} \\
    r_{01,10,01}  & \cdots & r_{01,10,22} & \cdots & r_{11,10,22} \\
    r_{01,11,01}  & \cdots & r_{01,11,22} & \cdots & r_{11,11,22} \\
  \end{array}
\right].
$$
$$
S^{2|3}=\left[
  \begin{array}{ccccc}
    v_{01,01}^{(23)} & v_{01,02}^{(23)} & v_{01,10}^{(23)}&\cdots& v_{01,22}^{(23)}\\
    v_{10,01}^{(23)} & v_{10,02}^{(23)} & v_{10,10}^{(23)} &\cdots& v_{10,22}^{(23)} \\
    v_{11,01}^{(23)} & v_{11,02}^{(23)} & v_{11,10}^{(23)} &\cdots& v_{11,22}^{(23)}\\
  \end{array}
\right],
$$

\begin{theorem}\label{1}
If the tripartite state $\rho\in H_1^{d_1}\otimes H_2^{d_2}\otimes H_3^{d_3}$ is separable under the bipartition $i|jk$, we have
\begin{equation}\label{8}
\begin{split}
&\|N^{i|jk}\|_{tr}\\
&\leq\sqrt{d_i-1}\left(|\alpha|\sqrt{d_j-1}  +|\beta|\sqrt{d_k-1}+|\gamma|\sqrt{\mathfrak{m}_{jk}} \right).
\end{split}
\end{equation}
\end{theorem}
{\it Proof}~~ Suppose the tripartite quantum state is separable under the bipartition $i|jk$, then
\begin{equation}\label{9}
\rho_{i|jk}=\sum_sp_s\rho_s^i\otimes \rho_s^{jk}, \quad 0<p_s\leq1, \sum_sp_s=1,
\end{equation}
where the factor states are
\begin{equation}\label{10}
\rho_s^i=\frac{1}{d_i}(I_{d_i}+T_s^{(i)}A_{(i)}), 
\end{equation}
\begin{equation}\label{11}
\begin{split}
\rho_s^{jk}=&\frac{1}{d_jd_k}(I_{d_j}\otimes I_{d_k}+T_s^{(j)}A_{(j)}\otimes I_{d_k}\\&+I_{d_j}\otimes T_s^{(k)}A_{(k)}
+T_s^{(jk)}(A_{(j)}\otimes A_{(k)}) )
\end{split}
\end{equation}
Therefore
\begin{equation}\label{12}
S^{i|j}=\sum_sp_sT_s^{(i)}(T_s^{(j)})^t,~~ S^{i|jk}=\sum_sp_sT_s^{(i)}(T_s^{(jk)})^t,
\end{equation}
where $t$ stands for transpose. Lemma 1 and 2 imply that
\begin{equation*}
\begin{split}
&\|N^{i|jk}\|_{tr}
\leq\sum_sp_s(|\alpha|\|T_s^{(i)}\|\|T_s^{(j)}\|+ |\beta|\|T_s^{(i)}\|\|T_s^{(k)}\|\\
& \qquad\qquad\qquad+|\gamma|\|T_s^{(i)}\|\|T_s^{(jk)}\|  )\\
&\leq\sqrt{d_i-1}\left(|\alpha|\sqrt{d_j-1}+ |\beta|\sqrt{d_k-1}+|\gamma|\sqrt{\mathfrak{m}_{jk}}\right),
\end{split}
\end{equation*}
where we have used the norm property
and $\||\alpha\rangle\langle\beta|\|_{tr}=\||\alpha\rangle\|\||\beta\rangle\|$
for vectors $|\alpha\rangle$ and $|\beta\rangle$.
\qed

A mixed state is called {\it genuine multipartite entangled} (GME) if it cannot be expressed as a convex combination of biseparable states.
In this note we exclusively consider GME of {\it symmetrically
coherent quantum systems}, which have the property that if the quantum state is
biseparable in one bipartition $i|jk$, then the system will also be biseparable in other bipartitions
and moreover when the quantum state is a convex sum of biseparable quantum states $\rho=\sum_ip_i\rho_i$, then the summands $\rho_i$ will
obey $N(\rho_i)\leq N(\rho)$ for any bipartitioned $N$.
This coherent condition is satisfied by
physically important quantum states, while the state displays visible invariance under permutation symmetry in the Hilbert space.

For a symmetrically coherent tripartite state $\rho$, let $T(\rho)=
\min\{\|N^{1|23}\|_{tr}, \|N^{2|13}\|_{tr}, \|N^{3|12}\|_{tr}\}$
and define
\begin{align*}
\mathcal{K}_1&=\textrm{Max}_{(ijk)}\{\sqrt{d_i-1}\left(|\alpha|\sqrt{d_j-1} +|\beta|\sqrt{d_k-1}\right. \\
&+\left. |\gamma|\sqrt{\mathfrak{m}_{jk}} \right)\},
\end{align*}
where the maximum is taken over all permutations $(ijk)$ of $(123)$. 
We stress that the maximum is needed to generate a lower bound, as a general state can be entangled in an unexpected bipartition.

We have the following test of GME.
\begin{theorem}\label{2}
A symmetrically coherent mixed quantum state $\rho\in H_1^{d_1}\otimes H_2^{d_2}\otimes H_3^{d_3}$ is genuine tripartite entangled if $T(\rho)>\mathcal{K}_1$.
\end{theorem}
{\it Proof}~ Suppose $\rho$ is a biseparable mixed state, then
$\rho=\rho_1+\rho_2+\rho_3$, where $\rho_i$ are respectively
separable states $\rho_1=\sum_ip_i\rho_i^1\otimes\rho_i^{23}, \rho_2=\sum_j r_j\rho_j^2\otimes\rho_j^{13}\, \rho_3=\sum_k s_k\rho_k^3\otimes\rho_k^{12}$ with 
$\sum_i p_i+\sum_j r_j+\sum_k s_k=1$. As our quantum state is symmetric,
using Theorem 1 and the coherent property $N^{i|jk}(\rho_i)\leq N^{i|jk}(\rho)$ it follows that
\begin{equation*}
\begin{split}
T(\rho)&=\min\{\sum_i p_i\|N^{1|23}(\rho)\|_{tr}, \sum_jr_j\|N^{2|13}(\rho)\|_{tr},\\
& \sum_ks_k\|N^{3|12}(\rho)\|_{tr}\}\leq\mathcal{K}_1.
\end{split}
\end{equation*}
Consequently, if $T(\rho)>\mathcal{K}_1$, $\rho$ is genuine tripartite entangled.
\qed

To see how our test fares, we consider the following example.
\begin{example} Let $\rho$ be the following Werner-type state over $H_1^3\otimes H_2^3\otimes H_3^2$:
\begin{equation}\label{14}
\rho=\frac{1-x}{18}I_{18}+x|\varphi\rangle\langle\varphi|,
\end{equation}
where $|\varphi\rangle=\frac{1}{\sqrt{5}}[(|10\rangle+|21\rangle)|0\rangle
+(|00\rangle+|11\rangle+|22\rangle)|1\rangle]$, $0\leq x\leq1$, and $I_{18}$ is the identity matrix. We can apply Theorem 1 to locate a
range of $x$ where $\rho$ is GME. Specifically, {let $\alpha=0, \beta=0,\gamma=1$, then Theorem 1 implies that when $0.51<x\leq1$, $\rho$ is GME,
which is large than the range $0.53\leq x\leq1$ given in
\cite{zljw}} (here we have corrected the theorem therein). We have compiled a comparison of several choices of the parameters $\alpha, \beta, \gamma$
in Table \ref{tab:2}, which clearly shows that our criterion detects more GME region than that of \cite{zljw}.

In another situation, {let $\alpha=1$, $\beta=0.03$, $\gamma = 1$, $\rho$ is GME when $0.81<x\leq1$, which is better than
 the interval $0.82\leq x\leq1$ given in \cite{zljw} for $\alpha = 1, \beta= 1$ given in \cite{zljw}} (here
 the latter $\beta$ corresponds to our $\gamma$).  

\begin{table}[!htb]
\caption{Comparison of GME regions using Thm. 1.}
\label{tab:2}
\centering
\begin{tabular}{cccc}
\hline\noalign{\smallskip}
\  & Range fr. \cite{zljw} & & Our range\\
\noalign{\smallskip}\hline\noalign{\smallskip}
$\alpha=\frac{1}{2}, \gamma=1$ & $0.69<x\leq 1$ & & $0.67<x\leq1$ \\
$\alpha=\frac{1}{3}, \gamma=2$ & $0.59<x\leq 1$ & & $0.57 <x\leq1$\\
$\alpha=0, \gamma=1$ & $0.53<x\leq 1$ & & $0.51 <x\leq1$\\
\noalign{\smallskip}\hline
\end{tabular}
\end{table}

\end{example}

\section{Genuine entanglement for multipartite quantum state}
 In this section we study GME for multiparite quantum states.
 The same matrix notation will be adopted. 

 Consider the tensor space $\otimes_{i=1}^nH_{d_i}$.
 Let $A^{(i)}=(A_{a_ib_j}^{(i)})$ be the column vector of Weyl operators $A_{a_ib_j}^{(i)}$ which acts on the $i$th factor of $\otimes_{i=1}^nH_{d_i}$ and identity on the other factors. Then any quantum state
$\rho$ over $\otimes_{i=1}^nH_{d_i}$ can be uniquely written as
\begin{equation}\label{15}
\begin{split}
\rho=&\frac{1}{d}(I_{d}+\sum_{i=1}^n T^{(i)}\cdot A_{(i)}+\sum_{1\leq i<j\leq n}T^{(ij)}\cdot A_{(i)}A_{(j)}\\
&\quad+\cdots+T^{(12\cdots n)}\cdot A_{(1)}\cdots A_{(n)}),
\end{split}
\end{equation}
where $d=d_1\cdots d_n$,  $T^{(i)}$ is the vector with components $t_{a_ib_i}^{(1)}=tr(\rho (A_{a_ib_i}^{(i)})^{\dagger})$,
$T^{(ij)}$ is the vector with components $t_{a_ib_i,a_jb_j}^{(ij)}=tr(\rho (A_{a_ib_i}^{(i)}A_{a_jb_j}^{(j)})^{\dagger})$, \ldots,
$T^{(12\cdots n)}$ is the vector with components $t_{a_1b_1, \ldots, a_nb_n}^{(12\cdots n)}=tr(\rho (A_{a_1b_1}^{(1)}\cdots A_{a_nb_n}^{(j)})^{\dagger})$.
Then we have
\begin{equation*}
\|T^{(i_1\cdots i_s)}\|^2=\sum_{(a_{j},b_{j})\neq(0,0)}|t_{a_{1}b_{1},\ldots, a_{s}b_{s}}^{(i_1\cdots i_s)}|^{2}.
\end{equation*}
Let $A_j$ be the sum of $\|T^{i_1\cdots i_j}\|^2$, i.e.
$A_1=\sum_{i=1}^n\|T^{(i)}\|^2,$
$A_2=\sum_{i<j}^n\|T^{(ij)}\|^2$, and 
\begin{equation}
A_s=\sum_{i_1<\cdots <i_s}\|T^{(i_1\cdots i_s)}\|^2.
\end{equation}
\begin{lemma}\label{3}
Let $\rho\in H_1^{d_1}\otimes\cdots\otimes H_n^{d_n}$ $(n>2)$ be an $n$-partite pure quantum state, $d=\max\{d_1,\cdots, d_n\}$ and $D=d_1\cdots d_n$. If $\frac{d_1d_2\cdots d_n}{d^2}\geq 1$, then
\begin{equation}\label{16}
\begin{split}
&\|T^{(12\cdots n)}\|^2\leq
\left(D-\frac{D}{n-1}\sum\limits_{s=1}^n\frac{1}{d_{s}^2}+\frac{1}{n-1}\right)\\
&+\frac{1}{n-2}\left(\frac{n}{n-1}-\frac{d_1\cdots d_n}{n-1}\sum\limits_{s=1}^n\frac{1}{d_{s}^2}\right).
\end{split}
\end{equation}
\end{lemma}
{\it Proof}~~For pure $\rho$, we have $tr(\rho^2)=1$ and $tr(\rho_{l_1}^2)=tr(\rho_{l_2\cdots l_n}^2)$ for any distinct indices $l_1, \ldots, l_n\in\{1,2,\cdots,n\}$. Here
$\rho_{l_1}$ and $\rho_{l_2\cdots l_n}$ are the reduced states for the subsystem $H_{l_1}^{d_{l_1}}$ and $H_{l_2}^{d_{l_2}}\otimes\cdots\otimes H_{l_n}^{d_{l_n}}$. Therefore, we have
\begin{equation}\label{17}
tr(\rho^2)=\frac{1}{d_1d_2\cdots d_n}(1+A_1+\cdots+A_n)=1,
\end{equation}
and
\begin{equation*}
\frac{1}{d_{l_1}}(1+\|T^{(l_1)}\|^2)=\frac{1}{D}(1+\|T^{(l_2)}\|^2+\cdots+\|T^{(l_2\cdots l_n)}\|^2).
\end{equation*}
Since $\sum\limits_{l_1=1}^n\frac{1}{d_{l_1}}tr(\rho_{l_1}^2)=\sum\limits_{l_1=1}^n\frac{1}{d_{l_1}}tr(\rho_{l_2\cdots l_n}^2)$,
we get that
\begin{equation*}
\sum\limits_{l_1=1}^n\frac{1}{d_{l_1}^2}(1+\|T^{(l_1)}\|^2)=\frac{1}{D}[n+(n-1)A_1
+\cdots+A_{n-1}].
\end{equation*}
Therefore,
\begin{equation}\label{19}
\begin{split}
A_1&=\frac{D}{n-1}\sum\limits_{s=1}^n\frac{1}{d_{s}^2}(1+\|T^{(s)}\|^2)
-\frac{n}{n-1}-\frac{n-2}{n-1}A_2\\
&-\frac{n-3}{n-1}A_3-\cdots-\frac{1}{n-1}A_{n-1}.
\end{split}
\end{equation}
Substituting (\ref{19}) into (\ref{17}), we get
\begin{equation*}
\begin{split}
&A_n=D-1-\frac{1}{n-1}\left(D\sum\limits_{s=1}^n\frac{1}{d_{s}^2}(1+\|T^{(s)}\|^2)
-n\right)\\
&-\frac{1}{n-1}A_2-\frac{2}{n-1}A_3-\cdots\frac{n-2}{n-1}A_{n-1}\\
&\leq\left(D-\frac{1}{n-1}D\sum\limits_{s=1}^n\frac{1}{d_{s}^2}+\frac{1}{n-1}\right)-\frac{1}{n-1}\frac{D}{d^2}A_1\\
&-\frac{1}{n-1}A_2
-\frac{2}{n-1}A_3-\cdots\frac{n-2}{n-1}A_{n-1}\\
&\leq\left(D-\frac{1}{n-1}D\sum\limits_{s=1}^n\frac{1}{d_{s}^2}+\frac{1}{n-1}\right)
-\frac{\left(A_1+\cdots+A_{n-1}\right)}{n-1}
\end{split}
\end{equation*}
Since $A_1+\cdots+A_{n-1}=D-1-A_n$, we have
\begin{equation}
\begin{split}
&A_n\leq \left(D-\frac{1}{n-1}D\sum\limits_{s=1}^n\frac{1}{d_{s}^2}+\frac{1}{n-1}\right)\\
&+\frac{1}{n-2}\left(\frac{n}{n-1}-\frac{D}{n-1}\sum\limits_{s=1}^n\frac{1}{d_{s}^2}\right)
\end{split}
\end{equation}
\qed

\begin{remark} The technical assumption $D\geq d^2$ is easily satisfied as we are considering multipartite case. It really means that
$d\leq (\prod_id_i)/d$, which usually holds in general.
\end{remark}
\begin{remark} Note that the first term of the bound is exactly that of \cite{zljw}.
For the case $D\geq {d^2}$, the second term $\frac{n}{n-1}-\frac{D}{n-1}\sum\limits_{s=1}^n\frac{1}{d_{s}^2}\leq 0$.
So the lower bound of $\|T^{(12\cdots n)}\|^2$ is sharper than that of \cite{zljw} in this case.
Actually in general, our bound is almost always significantly stronger than that of \cite{zljw}.
For instance when $d_i=d$, the second term in the bound is less than $-\frac{n}{(n-1)(n-2)}(d^{n-2}-1)$, which means that
the new bound is significantly sharper.
\end{remark}

Let $\rho$ be a general symmetrically coherent $n$-partite state $\rho\in H_1^{d_1}\otimes H_2^{d_2}\otimes\cdots\otimes H_n^{d_n}$ represented as \eqref{15}. For real numbers $\alpha$, $\beta$ and distinct
indices $l_1, \ldots, l_n\in\{1, 2, \cdots, n\}$,
we define the following block matrix
\begin{equation}\label{21}
N^{l_1\cdots l_{k-1}|l_k\cdots l_{n}}=\alpha S_0^{l_1\cdots \l_{k-1}|l_k}+\beta S^{l_1\cdots l_{k-1}|l_k\cdots l_n},
\end{equation}
for $k-1=1, 2,\cdots, [n/2]$, the smallest integer less or equal to $n/2$. Here
$S_0^{l_1\cdots l_{k-1}|l_k}$ is the block matrix
$S_0^{l_1\cdots \l_{k-1}|\l_k}=[S^{l_1\cdots \l_{k-1}|\l_k}~~O_{l_1\cdots l_{k-1}}]$, where
$S^{l_1\cdots \l_{k-1}|\l_k}=[t_{i_{l_1}j_{l_1}\cdots i_{l_k}j_{l_k}}^{(l_1\cdots l_k)}]$ is a $\prod\limits_{s=1}^{k-1}(d_{l_s}^2-1)\times
(d_{l_k}^2-1)$ matrix and $O_{l_1\cdots l_{k-1}}$ is the $\prod\limits_{s=1}^{k-1}(d_{l_s}^2-1)\times
[\prod\limits_{s=k}^{n}(d_{l_s}^2-1)-(d_{l_k}^2-1)]$
zero matrix. $S^{l_1\cdots l_{k-1}|l_k\cdots l_n}=[t_{i_1j_1,\cdots, i_nj_n}]$ is a $\prod\limits_{s=1}^{k-1}(d_{l_s}^2-1)\times
\prod\limits_{s=k}^{n}(d_{l_s}^2-1)$ matrix. For example, when $\rho\in H_1^{2}\otimes H_2^{2}\otimes H_3^{2}\otimes H_4^{3}$, $N^{13|24}=\alpha S_0^{13|2}+\beta S^{13|24}$, where
$$S^{13|2}=\left[
           \begin{array}{ccc}
             t_{01,01,01}^{(132)} & t_{01,10,01}^{(132)} & t_{01,11,01}^{(132)} \\
             t_{01,01,10}^{(132)} & t_{01,10,10}^{(132)} & t_{01,11,10}^{(132)} \\
             t_{01,01,11}^{(132)} & t_{01,10,11}^{(132)} & t_{01,11,11}^{(132)} \\
             \vdots & \vdots & \vdots \\
             t_{11,01,11}^{(132)} & t_{11,10,11}^{(132)} & t_{11,11,11}^{(132)} \\
           \end{array}
         \right],$$
$$S^{13|24}=\left[
           \begin{array}{ccccc}
             t_{01,01,01,01} & t_{01,01,01,02} & \cdots & t_{01,11,01,22}\\
             t_{01,01,10,01} & t_{01,01,10,02} & \cdots & t_{01,11,10,22}\\
             t_{01,01,11,01} & t_{01,01,11,02} & \cdots & \cdot\\
             \vdots & \vdots & \vdots  & \vdots   \\
             t_{11,01,11,01} & t_{11,01,11,02} & \cdots  & \cdot  \\
           \end{array}
         \right].
$$

Set $\mathfrak{n}_{l_1l_2\cdots l_k}=\left(d_{l_1}\cdots d_{l_k}-\frac{d_{l_1}\cdots d_{l_k}}{k-1}\sum\limits_{s=1}^k\frac{1}{d_{l_s}^2}+\frac{1}{k-1}\right)
+\frac{1}{k-2}\left(\frac{k}{k-1}-\frac{d_{l_1}\cdots d_{l_k}}{k-1}\sum\limits_{s=1}^k\frac{1}{d_{l_s}^2}\right)
$ for $k\geq 3$ and $\mathfrak{n}_{l_1l_2}=\mathfrak{m}_{l_1l_2}$.
\begin{theorem}\label{th3}
If the $n$-partite quantum state $\rho\in H_1^{d_1}\otimes H_2^{d_2}\otimes\cdots\otimes H_n^{d_n}$ is separable under the bipartition $l_1\cdots l_{k-1}|l_k\cdots l_{n}$ and $\frac{d_{l_1}\cdots d_{l_{k-1}}}{d_{l_i}^2}\geq 1$, $\frac{d_{l_k}\cdots d_{l_{n}}}{d_{l_j}^2}\geq 1$ for $i=1,\cdots k-1$ and $j=k,\cdots n$
then we have that\\
(i) $\|N^{l_1|l_{2}\cdots l_{n}}\|_{tr}\leq \mathcal{M}_{l_1}$;\\
(ii) $\|N^{l_1\cdots l_{k-1}|l_k\cdots l_{n}}\|_{tr}\leq \mathcal{M}_{l_1\cdots l_{k-1}}$ $(k\geq3)$;\\
where $$\mathcal{M}_{l_1}=\sqrt{d_{l_1}-1}\left(|\alpha|\sqrt{d_{l_2}-1}+|\beta|
\sqrt{\mathfrak{n}_{l_2\cdots l_n}}\right),$$
$$\mathcal{M}_{l_1\cdots l_{k-1}}=\sqrt{\mathfrak{n}_{l_1\cdots l_{k-1}}}
\left(|\alpha|\sqrt{d_{l_k}-1}+|\beta|\sqrt{\mathfrak{n}_{l_k\cdots l_n}}\right).$$
\end{theorem}
{\it Proof}~~(i) If the $n$-partite mixed state is separable under the bipartition $l_1|l_2\cdots l_{n}$, it can be expressed as
\begin{equation}\label{22}
\rho_{l_1|l_2\cdots l_{n}}=\sum\limits_s p_s\rho_s^{l_1}\otimes\rho_s^{l_2\cdots l_{n}}, \ 0<p_s\leq1, \sum\limits_s p_s=1,
\end{equation}
where
\begin{align}\label{23}
\rho_s^{l_1}&=\frac{1}{d_{l_1}}(I_{d_{l_1}}+T_s^{(l_1)}A_{(l_1)})\\ \notag
\rho_s^{l_2\cdots l_{n}}&=\frac{d_{l_1}}{D}(I_{D/d_{l_1}}+\sum_{i=2}^n T_s^{(l_{i})}\cdot A_{(l_{i})}+\cdots\\
&\qquad+T_s^{(l_2\cdots l_n)}\cdot A_{(l_2)}\cdots A_{(l_n)}) \label{24}
\end{align}
Then,
\begin{equation}\label{25}
\begin{split}
S^{l_1|l_2}
&=\sum\limits_sp_sT_s^{(l_1)}(T_s^{(l_2)})^t, \\
S^{l_1|l_2\cdots l_n}&=\sum\limits_sp_sT_s^{(l_1)}(T_s^{(l_2\cdots l_n)})^t.
\end{split}
\end{equation}
By Lemma 1 and Lemma 3, we have
\begin{equation*}
\begin{split}
&\|N^{l_1|l_2\cdots l_{n}}\|_{tr}\\
&\leq\sum_sp_s(|\alpha|\|T_s^{(l_1)}\|\|T_s^{(l_2)}\|+|\beta|\|T_s^{(l_1)}\|\|T_s^{(l_2\cdots l_n)}\|)\\
&\leq\sqrt{d_{l_1}-1}\left(|\alpha|\sqrt{d_{l_2}-1}+|\beta|
\sqrt{\mathfrak{n}_{l_2\cdots l_n}}\right)=\mathcal{M}_{l_1}.
\end{split}
\end{equation*}
(ii) If $\rho$ is separable under the bipartition $l_1\cdots l_{k-1}|l_k\cdots l_{n}$, then it is written as
\begin{equation}\label{26}
\rho_{l_1\cdots l_{k-1}|l_k\cdots l_{n}}=\sum\limits_s p_s\rho_s^{l_1\cdots l_{k-1}}\otimes\rho_s^{l_k\cdots l_{n}},
\end{equation}
where $0<p_s\leq1, \sum\limits_s p_s=1$.
\begin{equation}\label{27}
\begin{split}
&\rho_s^{l_1\cdots l_{k-1}}=\frac{1}{d_{l_1}\cdots d_{l_{n}}}(I_{d_{l_1}\cdots d_{l_{n}}}+\sum_{i=1}^{k-1} T_s^{(l_{i})}\cdot A_{(l_{i})}\\
&+\sum_{1\leq i<j\leq k-1}T_s^{(l_il_j)}\cdot A_{(l_i)}A_{(l_j)}+\cdots+\\
&+T_s^{(l_1\cdots l_{k-1})}\cdot A_{(l_1)}\cdots A_{(l_{k-1})})\\
\end{split}
\end{equation}
\begin{equation}\label{28}
\begin{split}
&\rho_s^{l_k\cdots l_{n}}=\frac{1}{d_{l_k}\cdots d_{l_{k-1}}}(I_{d_{l_k}\cdots d_{l_{n}}}+\sum_{i=k}^{n} T_s^{(l_{i})}\cdot A_{(l_{i})}\\
&+\sum_{k\leq i<j\leq n}T_s^{(l_il_j)}\cdot A_{(l_i)}A_{(l_j)}+\cdots)
\end{split}
\end{equation}
By definition it follows that
\begin{equation}\label{29}
\begin{split}
S^{l_1\cdots l_{k-1}|l_k}&=\sum\limits_sp_sT_s^{(l_1\cdots \l_{k-1})}(T_s^{(l_k)})^t,\\
S^{l_1\cdots l_{k-1}|l_k\cdots l_n}&=\sum\limits_sp_sT_s^{(l_1\cdots \l_{k-1})}(T_s^{(l_k\cdots l_n)})^t.
\end{split}
\end{equation}
Therefore
\begin{equation*}
\begin{split}
&\|N^{l_1\cdots l_{k-1}|l_k\cdots l_{n}}\|_{tr}\\
\leq&\sum_sp_s(|\alpha|\|T_s^{(l_1\cdots l_{k-1})}\|\|T_s^{(l_k)}\|\\
 & \qquad+|\beta|\|T_s^{(l_1\cdots l_{k-1})}\|\|T_s^{(l_k\cdots l_n)}\|)\\
\leq&\sqrt{\mathfrak{n}_{l_1\cdots l_{k-1}}}
(|\alpha|\sqrt{d_{l_k}-1}+|\beta|\sqrt{\mathfrak{n}_{l_k\cdots l_{n}}})\\
=&\mathcal{M}_{l_1\cdots l_{k-1}}.
\end{split}
\end{equation*}
\qed

Now we let's study genuine multipartite entanglement for symmetrically coherent quantum states. An $n$-partite mixed state $\rho=\sum p_i|\varphi_i\rangle\langle\varphi_i|$ is called {\it biseparable} if $|\varphi_i\rangle$ $(i=1,2,\cdots,n)$ can be expressed as one of the forms: $|\varphi_i\rangle=|\varphi_i^{l_1\cdots\l_{k-1}}\rangle\otimes|\varphi_i^{l_k\cdots l_n}\rangle$, where $|\varphi_i^{l_1\cdots\l_{k-1}}\rangle$ and $|\varphi_i^{l_k\cdots l_n}\rangle$ are some pure states in $H_{l_1}^{d_{l_1}}\otimes\cdots \otimes H_{l_{k-1}}^{d_{l_{k-1}}}$ and $H_{l_k}^{d_{l_k}}\otimes\cdots \otimes H_{l_{n}}^{d_{l_{n}}}$, respectively, $l_1\neq \cdots\neq l_n\in\{1,2,\cdots,n\}$. Otherwise, $\rho$ is said to be genuinely multipartite entangled. A mixed state is biseparable if it is a convex sum of biseparable pure states. Let $m=[n/2]$ and
\begin{equation}\label{30}
\begin{split}
&T(\rho)=\min\{\|N^{l_1|l_2\cdots l_n}\|_{tr}, \|N^{l_1l_2|l_3\cdots l_n}\|_{tr}, \cdots,\\
&\qquad \|N^{l_1\cdots l_m|l_{m+1}\cdots l_n}\|_{tr}\},
\end{split}
\end{equation}
where $l_1=1,2,\cdots,n$, $l_1<\cdots<\l_{k-1}\in\{1,2,\cdots,n\}$.
Similarly set $\mathcal{K}_2=\max\{\mathcal{M}_{l_1},\cdots, M_{l_1\cdots l_{[n/2]}}\}$,
where $l_1, \ldots, l_n$ as before.
We have the following criterion.
\begin{theorem}\label{4}
Assume that
$\frac{d_{l_1}\cdots d_{l_{k-1}}}{d_{l_i}^2}\geq 1$, $\frac{d_{l_k}\cdots d_{l_{n}}}{d_{l_j}^2}\geq 1$ for $i=1,\cdots k-1$ and $j=k,\cdots n$,
a mixed symmetrically coherent quantum state $\rho\in H_1^{d_1}\otimes H_2^{d_2}\otimes\cdots\otimes H_n^{d_n}$ is genuine multipartite entangled if $T(\rho)>\mathcal{K}_2$.
\end{theorem}
{\it Proof}~~Suppose $\rho$ is biseparable, it follows from Theorem 3 and \eqref{30} that
\begin{equation*}
T(\rho)\leq \mathcal{K}_2
\end{equation*}
Consequently, if $T(\rho)>\mathcal{K}_2$, $\rho$ is genuine multipartite entangled.
\qed

Suppose the Hilbert space is $(H^d)^{\otimes n}$. This includes the case of symmetric density matrix.
similar to Corollary 1, the following result is easily established. 

\begin{corollary} 
 Let $\rho$ be a symmetric density matrix over $(H^d)^{\otimes n}$ ($n\geq 4$). Set $m=[n/2]$. Then $\rho$ is genuine multipartite entangled
if $T(\rho)>\mathcal{J}_2$, where
$$\mathcal{J}_2=\max\{\mathcal{M}_{(1)}, \mathcal{M}_{(2)}, \cdots, \mathcal{M}_{(m)}\},$$
and $\mathcal M_{(k)}=\sqrt{d^k-\frac{k}{k-2}d^{k-2}+\frac2{k-2}}(|\alpha|\sqrt{d-1}+|\beta|\sqrt{d^{n-k}-\frac{n-k}{n-k-2}d^{n-k-2}+\frac2{n-k-2}})$
\end{corollary}

\begin{figure}[!htb]
  \centering
  \includegraphics[scale=0.5]{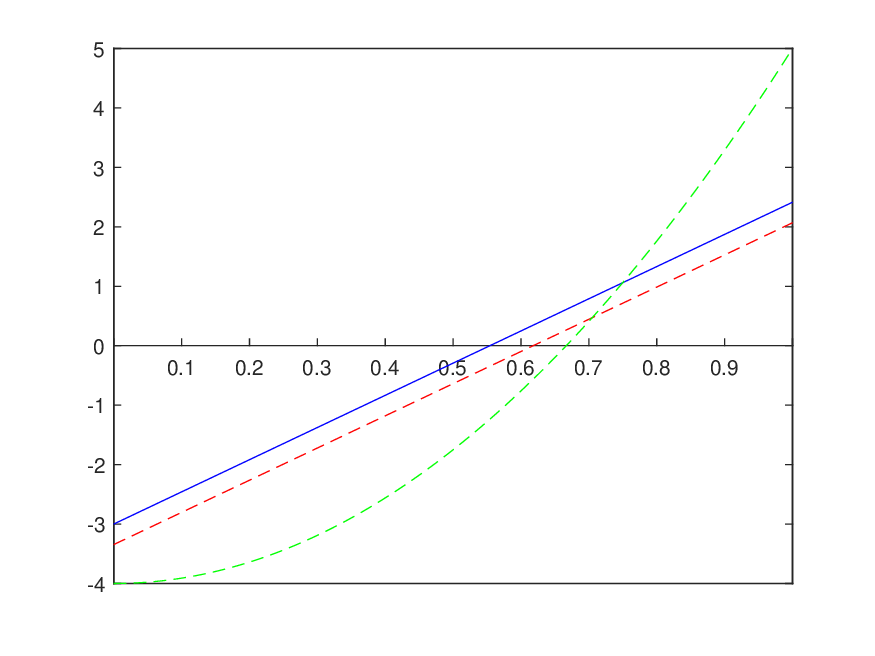}
  \caption{$F_1(x)$ from our result (solid straight line), $G_1(x)$ from \cite[Th.3]{zljw} (red dashed straight line), $G_2(x)$ from
   \cite[Th.3]{lww} (green dashed curve line ) }
  \label{fig:2}
\end{figure}

\textit{\textbf{Example 2}} Consider the four-qubit state $\rho\in H_1^2\otimes H_2^2\otimes H_3^2\otimes H_4^2$,
\begin{equation}\label{31}
 \rho=x|\psi\rangle\langle\psi|+\frac{1-x}{16}I_{16},
\end{equation}
where $|\psi\rangle=\frac{1}{\sqrt{2}}(|0000\rangle+|1111\rangle)$, $0\leq x\leq1$, and
$I_{16}$ is the identity matrix in $M_{16}(\mathbb C)$.
Using Theorem 3 (i) for the bipartition $l_1|l_2l_3l_4$ with $\alpha=1, \beta=1$, we set
$F_1(x)=\|N^{l_1|l_2l_3l_4}\|_{tr}-3=(4+\sqrt{2})x-3$. Then $\rho$ is not separable under the bipartition for $F_1(x)>0$, i.e. $0.5541<x\leq1$.
However by \cite[Th.3]{zljw}, $\rho$ is not separable under the bipartition for $G_1(x)=(4+\sqrt{2})x-(1+\sqrt{11/2})>0$, i.e. $0.6179<x\leq1$.
Similarly using \cite[Th.3]{lww}, $\rho$ is not separable under the bipartition for $G_2(x)=9x^2-4$, i.e. $0.6667<x\leq1$. Fig. \ref{fig:2} shows that our method is stronger compared with some of the recent tests. 

Now using Corollary 2 and set $\alpha=1, \beta=1$, we get that $F_2(x)=T(\rho)-J_2=5x-4.7321$,
$\rho$ is genuine tripartite entanglement for $F_2(x)>0$, i.e. $0.94<x\leq1$, our range is the same as the range according to \cite[Cor. 2]{zljw}.

\section{Conclusions}
Using the generalized Bloch representations of density matrices via Weyl operators, we have come up
with several general tests to determine genuine entanglement for multipartite quantum systems.
Our approach starts with some finer upper bounds for the norms of the correlation tensors, which then lead to then new entanglement criteria for genuine tripartite entangled quantum states.
The key technical point is based on certain matrices compiled from the subtensors of the
correlation tensor of the density matrices.
The results are then generalized to higher dimensional multipartite symmetrically coherent quantum systems to detect genuine entanglement in arbitrary dimensional quantum states.
Compared with previously available criteria, our new results detect more situations, which are explained
in details with several examples. When the number of the particles is large, our criteria is found to be significantly stronger to detect GME.

\bigskip

\textbf{Acknowledgments}

This work is supported by the National Natural Science Foundation of China under
grant no.  12171303  
and Simons Foundation under grant no. 523868.



%

%
%


\end{document}